\documentclass[12pt]{iopart}
 \usepackage{graphicx}
 \usepackage{float}
 \usepackage[numbers]{natbib}
 \usepackage{hyperref}
 
\begin{document}
\title[Unified Wronskian formulation of inverse scattering with SUSYQM]
{Unified Wronskian formulation of inverse scattering with supersymmetric quantum mechanics}
\author{Q Bozet$^{1,2}$ and J-M Sparenberg$^2$}
\address{$^1$ Institut f\"ur Kernphysik, Johannes Gutenberg-Universit\"at
Mainz, D-55099 Mainz, Germany}
\address{$^2$ Nuclear Physics and Quantum Physics,
École polytechnique de Bruxelles,
Université libre de Bruxelles (ULB) - CP 229, B-1050 Brussels, Belgium}
\ead{qbozet@uni-mainz.de $\mathrm{and}$ jmspar@ulb.ac.be}

\begin{abstract}

The Wronskian formulation of supersymmetric quantum mechanics (SUSYQM) confluent transformation pairs \cite{bermudez2012,schulze2013} is applied to the construction of phase-equivalent potentials with different bound spectra, replacing integral formulas.
This allows to unify the two steps of a SUSYQM inversion scheme consisting in (i) the construction of a unique bound-state-less potential, possibly singular, from phase-shift inversion by a chain of non-confluent SUSYQM transformations, and (ii) the phase-equivalent addition of bound states by confluent SUSYQM pairs.
Both steps are now combined in a single Wronskian formula, providing an elegant complete solution to the fixed-angular-momentum inversion problem.
This formalism is applied to the inversion of $^3S_1$ and $^1S_0$ neutron-proton data and its numerical implementation is discussed.

\end{abstract}
\submitto{\jpa}
\maketitle

\section{Introduction}

The problem addressed in this article is the two-body non-relativistic fixed-angular-momentum quantum inverse scattering \cite{chadan2012}. The objective is to reconstruct, for a given partial wave, an interaction potential between two particles, starting from experimental data, namely the scattering phase shift and the bound-state properties \cite{taylor1972}.
The traditional ways to solve this inverse problem are the Gel'fand-Levitan and Marchenko integral equations. Here, we follow a different approach, called the supersymmetric inversion method. It is based on supersymmetric quantum mechanics, an algebraic technique developed by Edward Witten in \cite{witten1981} as a non-relativistic version of relativistic supersymmetric field theories. A complete introduction to supersymmetric quantum mechanics can be found in \cite{junker2012}.

This inversion method was first developed in \cite{baye1987a,baye1987b,baye1994,sparenberg1997} and inspired by ideas laid out in \cite{sukumar1985a,sukumar1985b}. The general idea of the method is to first perform a series of supersymmetric transformations on an original potential, generally chosen to be purely centrifugal, in order to fix its phase shift and to make it comply with experimental scattering data. The supersymmetric transformations used in this first part of the inversion scheme are performed at pairwise different energies and are therefore called non-confluent. An analytical formulation, using Wronskians, was already found in \cite{crum1955}. It is called the Crum-Krein formula. The potential obtained by these non-confluent transformations is unique: it does not possess any bound state, which introduce ambiguities in the scattering inverse problem.
The price to pay for this uniqueness is that this potential is possibly singular at the origin.
This problem is thus called the singular inverse problem.

In order to construct a potential that enables reproducing both scattering-state and bound-state experimental data, bound states have to be added to this singular potential, without altering its phase shift. It is possible to do so by performing particular pairs of supersymmetric transformations, at the same energy. These are called phase-equivalent transformations and were first developed in \cite{baye1987a,baye1987b}. Reviews of the supersymmetric inversion method can be found in \cite{baye2004,baye2014}.

Until now, it proved impossible to obtain an analytical formulation of the final potential, due to the confluent character of the phase-equivalent transformations, as the Crum-Krein formula only applies to non-confluent transformations. The final transformations were therefore performed numerically thanks to integrals.

More recently, a new field has emerged, that will enable us to obtain a Wronskian formulation of the transformed potential and to combine the two separate steps of the inversion procedure, specifically the determination of the phase shift and the addition of the bound states. This field is called confluent supersymmetric quantum mechanics. The final analytical formula will resemble closely the Crum-Krein formula. Its aim is to study specifically confluent supersymmetric transformations. Seminal papers include \cite{bermudez2012,schulze2013}. These form the cornerstone of the present article.

Below, we first briefly review non-confluent supersymmetric transformations and the singular inverse problem (section 2),
then express phase-equivalent confluent transformations in terms of Wronskians, deriving the main new formula of the work (section 3), and finally we apply the formalism to the construction of neutron-proton potentials (section 4).

Throughout the work we use the reduced units $\hbar^2/2\mu=1$, where $\hbar$ is the reduced Planck constant and $\mu$ is the reduced mass of the two interacting particles.
\section{Supersymmetric quantum mechanics and
singular inverse problem}
Before presenting the main results of our work, we briefly review the supersymmetric inversion scheme.
We start from the radial Schrödinger equation at energy $E$,
\begin{equation}
    H_{0} \psi_{0}(r,E) = E \psi_{0}(r,E),
    \label{Schr0}
\end{equation}
where the Hamiltonian is given by
\begin{equation}
    H_{0} = - \frac{\mathrm{d}^{2}}{\mathrm{d}r^{2}} + V_{0}(r)
\end{equation}
and $V_0(r)$ is the effective interaction potential between the two particles separated by distance $r$.
The potential $V_0$ is assumed to decrease fast enough at large distances for solutions $\psi_0$ to behave like simple exponentials or trigonometric functions, and to satisfy the following behaviour at the origin,
\begin{equation}
    V_{0} \mathop{\rightarrow}_{r {\rightarrow}0} n_{0}(n_{0} + 1)r^{-2},
\end{equation}
where the positive integer $n_{0}$ is called the singularity parameter and may be larger than or equal to the orbital angular momentum quantum number $\ell$.
For positive energies $E=k^2$, where $k$ is the wave number, the regular solution at the origin satisfies
\begin{equation}
 \psi_0(r,E) \mathop{\sim}_{r\rightarrow 0} r^{n_0+1}
\end{equation}
and behaves asymptotically as
\begin{equation}
 \psi_0(r,E)  \mathop{\sim}_{r\rightarrow \infty} \sin[kr-\ell \textstyle\frac{\pi}{2}+\delta_0(k)],
\end{equation}
which defines the scattering phase shift $\delta_0(k)$ up to a multiple of $\pi$.

We then define the Hamiltonian
\begin{equation}
    H_{-} = H_{0} - \mathcal{E},
\end{equation}
where $\mathcal{E}$ is a constant called the factorization energy, generally assumed to be real negative. It is also noted $\mathcal{E} = - \kappa^{2}$, where $\kappa$ is a pseudo wave number.
If we suppose that $H_{-}$ is the negative superpartner of a supersymmetric quantum system, we can write (see \cite{junker2012} for more details)
\begin{equation}
    H_{0} = L L^{\dagger} + \mathcal{E},
\end{equation}
where 
\begin{eqnarray}
    L = - \frac{\mathrm{d}}{\mathrm{d}r} + \frac{\mathrm{d}}{\mathrm{d}r} \ln\varphi_{0}(r,\mathcal{E}), \\
    L^{\dagger} = \frac{\mathrm{d}}{\mathrm{d}r} + \frac{\mathrm{d}}{\mathrm{d}r} \ln\varphi_{0}(r,\mathcal{E}),
\end{eqnarray}
and $\varphi_{0}$ is called the factorization solution. It is a solution of the Schrödinger equation \eref{Schr0} at the factorization energy $\cal E$. We can define, through the positive superpartner of the original Hamiltonian, the new Hamiltonian
\begin{equation}
    H_{1} = L L^{\dagger} + \mathcal{E}.
\end{equation}
This operation is called a supersymmetric transformation. The transformed potential reads
\begin{equation}
    V_{1}(r) = V_{0}(r) - 2 \frac{\mathrm{d}^{2}}{\mathrm{d}r^{2}}\ln\varphi_{0}(r,\mathcal{E}), 
\end{equation}
which shows that $\varphi_0$ should not have any node to avoid singularities in potential $V_1$.

Solutions of the $V_1$ Schrödinger equation read
\begin{equation}
    \psi_{1}(r, E) = \mathrm{W}[\psi_{0}(r, E), \varphi_{0}(r, \mathcal{E})] \varphi_{0}(r, \mathcal{E})^{-1},
    \label{psi1}
\end{equation}
where $\mathrm{W}[\psi,\varphi]\equiv\psi\varphi'-\varphi'\psi$ is the Wronskian of the two functions and prime denotes the derivative with respect to $r$.
For $E\ne \mathcal{E}$, equation \eref{psi1} provides two linearly independent solutions of the $V_1$ equation, starting from two linearly independent solutions of the $V_0$ equation.
For $E = \mathcal{E}$, the Wronskian in equation \eref{psi1} is either zero or constant and we only have one solution
\begin{equation}
    \psi_{1}(r, \mathcal{E}) = \varphi_{0}(r,\mathcal{E})^{-1}.
\end{equation}
Due to the structure of the Schrödinger equation, a linearly independent solution $\psi_1^\perp$, called orthogonal solution, can always be constructed from this particular solution.
It reads
\begin{eqnarray}
    \psi_{1}(r,\mathcal{E})^{\perp} & = & \psi_{1}(r,\mathcal{E}) \int \frac{\mathrm{d}t}{\psi^{2}_{1}(t,\mathcal{E})}
    \label{psi1perp}\\
    & = & \varphi_{0}(r,\mathcal{E})^{-1} \int \varphi_{0}(t,\mathcal{E})^2\mathrm{d}t,
    \label{psi1perp_phi0}
\end{eqnarray}
where the primitives contain an arbitrary constant or integration bound,
and it satisfies the relation
\begin{equation}
    \mathrm{W}[\psi_{1}(r,\mathcal{E}),\psi_{1}(r,\mathcal{E})^{\perp}] = 1.
\end{equation}
The effects of the possible transformations on the bound spectrum and on the phase shifts depend on the boundary behaviours of the factorization solution.
They are summarized in table \ref{transformations}.

We can of course perform a number $M$ of supersymmetric transformations at factorization energies $\mathcal{E}_1, \dots, \mathcal{E}_M$ on a given potential. In that case, the final potential is given by
\begin{equation}
    V_{M}(r) = V_{0}(r) - 2 \frac{\mathrm{d}^2}{\mathrm{d}r^2} \sum^{M}_{m=1}\ln \varphi_{m-1}(r, \mathcal{E}_{m}).
    \label{transformed_potential}
\end{equation}
Chains of transformations at the same energy are called confluent, whereas if the energies are pairwise different, it is called a non-confluent chain. A chain with factorization energies that are not pairwise different is called multi-confluent. Note that in equation \eref{transformed_potential}, the factorization solutions used are not calculated with respect to the original potential but with the transformed potential. It is therefore in general impossible to have an analytical formula for the transformed potential, except if the chain is non-confluent. In that case, we have the Crum-Krein formula \cite{crum1955},
\begin{equation}
    V_{M}(r) = V_{0}(r) - 2 \frac{\mathrm{d}^{2}}{\mathrm{d}r^{2}}\ln\mathrm{W}[\varphi_{0}(r,\mathcal{E}_{1}),...,\varphi_{0}(r,\mathcal{E}_{M})].
    \label{VM}
\end{equation}
The solutions for the transformed potential are given by
\begin{equation}
    \psi_M(r, E) = \frac{\mathrm{W}[\varphi_{0}(r,\mathcal{E}_{1}),...,\varphi_{0}(r,\mathcal{E}_{M}), \psi_{0}(r,E)]}{\mathrm{W}[\varphi_{0}(r,\mathcal{E}_{1}),...,\varphi_{0}(r,\mathcal{E}_{M})]}.
    \label{psiM}
\end{equation}

We can now detail the first step of the inversion procedure. First, we fix the phase shift $\delta_M(k)$, expressed as a sum of arctangent terms, thanks to a sufficient number of non-confluent transformations $T_{\mathrm{l}}$ and $T_{\mathrm{r}}$, with real negative (or possibly complex) factorization energies chosen to fit the experimental data.
The potential built in this fashion is unique, as
it does not possess any bound state,  but it may be singular when the phase shift is larger at zero energy than at infinite energy. Indeed, according to the generalized Levinson theorem\cite{swan1963}, we have, in general
\begin{eqnarray}
    \delta(0) & = & N \pi, \\
    \delta(\infty) & = & - (n-\ell) \textstyle\frac{\pi}{2},
    \label{Levinson}
\end{eqnarray}
where $N$ is the number of bound states, $n$ is the singularity parameter of the potential and $\ell$ is the partial wave.
Since the phase shift is only defined up to a multiple of $\pi$ from experimental data, it is always possible to replace a bound state by increasing the singularity parameter of two units.

\begin{table}
{
{\renewcommand{\arraystretch}{1.5}
\hskip-1.0cm \begin{tabular}{|c|c|c|c|c|c|c|} 
 \hline
  Name & $\lim_{r \rightarrow 0} \varphi_{0}(r)$ & $n_{0}$ & $n_{1}$ &  $\lim_{r \rightarrow \infty} \varphi_{0}(r)$ & effect on bound spectrum & $\delta_{1}(k) - \delta_{0}(k)$\\
 \hline
 $T_{\mathrm{l}}(\mathcal{E})$  & $r^{n_{0}+1}$  & $\geq 0$ & $n_{0} + 1$ & $e^{\kappa r}$ & none & $-\arctan(\frac{k}{\kappa})$\\
 \hline
 $T_{\mathrm{r}}(\mathcal{E})$  & $r^{-n_{0}}$  & $>0$ & $n_{0} - 1$ & $e^{-\kappa r}$ & none & $\arctan(\frac{k}{\kappa})$\\
 \hline
  $T_{\mathrm{rem}}(\mathcal{E})$  & $r^{n_{0}+1}$  & $\geq 0$ & $n_{0} + 1$ & $e^{-\kappa r}$ & removal of bound state & $-\pi+\arctan(\frac{k}{\kappa})$\\
 \hline
 $T_{\mathrm{add}}(\mathcal{E}, \beta)$  & $r^{-n_{0}}$  & $>0$ & $n_{0} - 1$ & $e^{\kappa r}$ & addition of bound state & $\pi-\arctan(\frac{k}{\kappa})$\\
 \hline
 
\end{tabular}
}
}
\vspace{1.5em}
\caption{Effects on the characteristics of the potential of the different types of supersymmetric transformations. Adapted from \cite{baye2014}}.
\label{transformations}
\end{table}

\section{Phase-equivalent potentials and
confluent supersymmetric transformations}
In order to construct a physically realistic potential, we still need to add bound states. According to table \ref{transformations}, this can be done without modifying the phase shift (except for an addition of $\pi$) using a transformation $T_{\mathrm{r}}$ followed by a transformation $T_{\mathrm{add}}$ performed at the same energy, chosen as the desired bound-state energy.
Let $\mathcal{E}$ be this energy and $\varphi_M(r,\mathcal{E})$ be the solution regular at infinity for potential $V_M$,
used as factorization solution for the $T_{\mathrm{r}}$ transformation.
The general solution for potential $V_{M+1}$ that is singular both at the origin and at infinity can be obtained from equation \eref{psi1perp_phi0}. It reads, up to an arbitrary multiplicative factor,
\begin{equation}
\varphi_{M+1}(r,\mathcal{E}) =
\varphi_M(r,\mathcal{E})^{-1} \left[\beta + \int_r^{+\infty} \varphi_M(t,\mathcal{E})^2\mathrm{d}t\right],
\end{equation}
where the integration constants have been chosen to make the integral converge and constant $\beta$ should be chosen strictly positive to avoid a node in the solution.
Potential $V_{M+2}$ can then be obtained from equation \eref{transformed_potential}
and reads
\begin{eqnarray}
     V_{M+2}(r) & = & V_M(r) - 2 \frac{\mathrm{d}^{2}}{\mathrm{d}r^{2}}\ln      \varphi_M(r,\mathcal{E}) - 2 \frac{\mathrm{d}^{2}}{\mathrm{d}r^{2}}\ln      \varphi_{M+1}(r,\mathcal{E})
     \nonumber\\
     & = & V_{M}(r) - 2 \frac{\mathrm{d}^{2}}{\mathrm{d}r^{2}}\ln\left[ \beta +\int^{ + \infty}_{r} \varphi_{M}(t, \mathcal{E})^{2} \mathrm{d}t\right],
     \label{integral_formula}
\end{eqnarray}
where the factorization solution of the first transformation has simplified, leading to an elegant integral formula.
The arbitrary choice of $\mathcal{E}$ and $\beta$ guarantees that all potentials with phase shift $\delta_M$ and one bound state are accessible through this formula, the choice of $\beta$ allowing to choose the asymptotic normalization constant (ANC) of the bound state.

Equation \eref{integral_formula} is actually a particular case of phase-equivalent transformation pair.
For a potential already displaying a bound state, we can also remove this bound state by using a transformation $T_{\mathrm{rem}}$ followed by a transformation $T_{\mathrm{l}}$ at the bound-state energy. Using a transformation $T_{\mathrm{rem}}$ followed by a transformation $T_{\mathrm{add}}$ does not modify the phase shift nor the bound spectrum of the potential but can modify the ANC of the bound state, again thanks to parameter $\beta$.
In both cases, the factorization solution $\varphi_M(r, \mathcal{E})$ should be chosen as the normalized bound-state wave function. There are thus three types of phase-equivalent transformation pairs, which can all be expressed in terms of the same formula \eref{integral_formula}, but with different values of $\beta$ and different types of factorization solutions \cite{baye2004}.
They are summarized in table \ref{pairs} for $M=0$.

Coming back to the complete inversion scheme, the difficulty of equation \eref{integral_formula} is that it cannot be combined with equation \eref{VM} in a unified form, where only solutions of the original potential $V_0$ would appear. Because of the complicated expression of $\varphi_M$, given by equation \eref{psiM}, the integral in equation \eref{integral_formula} is in general not known analytically, even if the original potential $V_{0}$ is chosen to be simple. The only way to perform phase-equivalent bound-state additions is thus purely numerical.
In light of the considerations of section 2, the logical next step would be to obtain an analytical formula for potential $V_{M+2}$ and its generalization to more bound states.

The results of \cite{bermudez2012,schulze2013} make this possible. Indeed, in \cite{schulze2013} a Wronskian formula was discovered to express the effects of a multi-confluent chain of arbitrary length on an original potential $V_{0}$.
In the present case, the general theorem proved in \cite{schulze2013} gives the following result for the final potential $V_{M+2}$
\begin{eqnarray}
    V_{M+2}(r) & = & V_{0}(r) - 2 \frac{\mathrm{d}^{2}}{\mathrm{d}r^{2}}\ln \mathrm{W}\left[\varphi_{0}(r,\mathcal{E}_{1}),..., \varphi_{0}(r,\mathcal{E}_{M}),\right. \nonumber \\
    && \left. u_{0}(r,\mathcal{E}_{M+1}), u_{1}(r,\mathcal{E}_{M+1})\right], \label{Wronskian}
\end{eqnarray}
where $\varphi_{0}(r,\mathcal{E}_{i}$) is a factorization solution calculated with the original potential at energy $\mathcal{E}_{i}$, as before,
and $u_{0}(r,\mathcal{E}_{{\mathrm{M+1}}})$ and $u_{1}(r,\mathcal{E}_{{\mathrm{M+1}}})$ are the generalized eigenfunctions of the following Jordan chain of length two
\begin{eqnarray}
    (H - \mathcal{E}_{\mathrm{M+1}} ) u_{0} = 0, \\
    (H - \mathcal{E}_{\mathrm{M+1}} ) u_{1} = u_{0}.
\end{eqnarray}
It can also be formulated in another way
\begin{eqnarray}
    (H - \mathcal{E}_{\mathrm{M+1}} ) u_{0} = 0, \\
    (H - \mathcal{E}_{\mathrm{M+1}} )^{2} u_{1} = 0.
\end{eqnarray}

An expression for the generalized eigenfunction $u_{1}$ was discovered in \cite{bermudez2012}. It reads, for $u_{0} = \varphi_{0}$,
\begin{equation}
    u_{1}(r,\mathcal{E}_\mathrm{M+1}) =
    \beta \varphi_{0}(r,\mathcal{E}_{\mathrm{M+1}})^{\perp} + \frac{\partial}{\partial \mathcal{E}} \varphi_{0}(r,\mathcal{E}_{\mathrm{M+1}}), \label{gen_eigenfunctions}
\end{equation}
where $\beta$ is an arbitrary constant and $\frac{\partial}{\partial \mathcal{E}} \varphi_{0}(r,\mathcal{E}_{\mathrm{M+1}})$ is the partial derivative of the factorization solution with respect to factorization energy evaluated at energy $\mathcal{E}_{\mathrm{M+1}}$.
The orthogonal solution $\varphi_{0}(r,\mathcal{E}_{\mathrm{M+1}})^{\perp}$ is defined, as in equation \eref{psi1perp}, by
\begin{equation}
    \varphi_{0}(r,\mathcal{E}_{\mathrm{M+1}})^{\perp} = \varphi_{0}(r,\mathcal{E}_{\mathrm{M+1}}) \int \frac{\mathrm{d}t}{\varphi^{2}_{0}(t,\mathcal{E}_{\mathrm{M+1}})}
\end{equation}
and contains an arbitrary constant or integration bound in the primitive.
Combining equations \eref{Wronskian} and \eref{gen_eigenfunctions}, we obtain the following formula,
\begin{eqnarray}
    V_{M+2}(r) & = & V_{0}(r) - 2 \frac{\mathrm{d}^{2}}{\mathrm{d}r^{2}}\ln \mathrm{W}\bigg[\varphi_{0}(r,\mathcal{E}_{1}),..., \varphi_{0}(r,\mathcal{E}_{M}), \nonumber \\
 && \varphi_{0}(r,\mathcal{E}_{M+1}),     \beta \varphi_{0}(r,\mathcal{E}_{\mathrm{M+1}})^{\perp} + \frac{\partial}{\partial \mathcal{E}} \varphi_{0}(r,\mathcal{E}_{\mathrm{M+1}})\bigg] . \label{final_formula}
\end{eqnarray}
This is the main result of the present paper: the final potential is now expressed in terms of factorization solutions calculated with the original potential $V_{0}$ only. If $V_0$ is simple enough, equation \eref{final_formula} is analytic. This formula can be extended to the case of the addition of an arbitrary number of bound states using the results of \cite{schulze2013},
simply by adding pairs of functions $u_0$ and $u_1$ in the Wronskian for different bound-state energies.

We can also derive the precise link between the two concurrent formulations, integral and differential, of the three types of phase-equivalent pairs. Comparing equations \eref{final_formula} and \eref{integral_formula} for the case of a single pair of confluent supersymmetric transformations, i.e.\ for $M=0$ with $\mathcal{E}_1=\mathcal{E}$,
and using the identity
\begin{equation}
    \mathrm{W}[u_{0}, u_{1}] = \beta + \mathrm{W}\left[u_{0}, \frac{\partial}{\partial \mathcal{E}} \varphi_{0}(r,\mathcal{E})\right],
\end{equation}
we see that parameter $\beta$ plays the same role in both expressions and we get the non-trivial relation
\begin{equation}
    \int^{ + \infty}_{r} \varphi_0(t, \mathcal{E})^2 \mathrm{d}t = \mathrm{W}\left[\varphi_{0}(r,\mathcal{E}), \frac{\partial}{\partial \mathcal{E}} \varphi_{0}(r,\mathcal{E})\right].
    \label{non-trivial}
\end{equation}
This identity can also be derived by using results from \cite{salinas2011} and \cite{bermudez2012}. Combining this equation with the parameters of  table \ref{pairs} thus provides a Wronskian formulation of the three phase-equivalent pairs.

\begin{table}
{
{\renewcommand{\arraystretch}{1.5}
\hskip-1.0cm \begin{tabular}{|c|c|c|c|c|c|c|} 
 \hline
  $\mathcal{E}$ & $\lim_{r \rightarrow 0} \varphi_{0}(r)$ & $\beta$ & $n_{2}$ &  $\lim_{r \rightarrow \infty} \varphi_{0}(r)$ & effect on bound spectrum & $\delta_{2}(k) - \delta_{0}(k)$\\
 \hline
  $=E_{0}$  & $r^{n_{0}+1}$  & $-1$ & $n_{0} + 2$ & $e^{-\kappa r}$ & removal of bound state & $-\pi$\\
 \hline
 $= E_{0}$  & $r^{n_{0}+1}$  & $< -1 \ \mathrm{or} > 0$ & $n_{0}$ & $e^{-\kappa r}$ & none & $0$\\
 \hline
 $\ne E_{0}$  & $r^{-n_{0}}$  & $>0$ & $n_{0} - 2$ & $e^{-\kappa r}$ & addition of bound state & $+\pi$\\
 \hline
\end{tabular}
}
}
\vspace{1.5em}
\caption{Effects of the different pairs of phase-equivalent supersymmetric transformations given by formula \eref{integral_formula} for $M=0$. $E_{0}$ represents the energy of a bound state. Adapted from \cite{baye2004}}
\label{pairs}
\end{table}
\section{Application to the neutron-proton system}

We are now going to apply formula \eref{final_formula} to two concrete examples: the inversion of the triplet and singlet neutron-proton $S$-wave phase shifts and the construction of the corresponding analytical potentials.

\subsection{$^{3}S_{1}$ state}
For the triplet state, we focus on low energies,
as for this state, the neutron-proton system possesses a well-known weakly bound state,
the deuteron (here we neglect the tensor term of the interaction and the $^3D_1$ wave component of the wave function).
Our aim is to build the most general potential fitting the low-energy phase shift and displaying one bound state with arbitrary energy and asymptotic normalization constant (ANC).

Following reference \cite{baye2014}, we start from the initial potential $V_0=0$, which has simple mathematical solutions for an $S$ wave at negative energies,
namely purely exponential or hyperbolic functions.
In that case, the supersymmetric inversion scheme leads to generalized Bargmann potentials.
The low-energy phase shift is well reproduced by the expression
\begin{equation}
    \delta_2(k) = -\arctan\frac{k}{\kappa_0}
    - \arctan\frac{k}{\kappa_1},
     \label{d2}
\end{equation}
where the factorization parameters, $\kappa_{0}$ and $\kappa_{1}$, are related to the scattering length
and effective range
and take the values $\kappa_{0} = 0.9090~\mathrm{fm}^{-1}$ and $\kappa_{1} = 0.2315~\mathrm{fm}^{-1}$.
According to table \ref{transformations},
the corresponding potential can be build through two non-confluent supersymmetric transformations,
with hyperbolic sine functions as factorization functions.
It reads, in reduced units,
\begin{equation}
    V_2(r) = -2 \frac{\mathrm{d}^{2}}{\mathrm{d}r^{2}}\ln \mathrm{W}[\sinh(\kappa_{0} r), \sinh(\kappa_{1} r)],
\end{equation}
and displays a $n_2=2$ singularity at the origin.

We then add a bound state at an arbitrary energy
${\cal E}_2 = -\frac{\hbar^2}{2\mu}\kappa_2^2$ with two confluent transformations.
In order to comply with the requirements of table \ref{pairs}, we choose a decreasing exponential function $\varphi_0(r,{\cal E}_2) = e^{-\kappa_2 r}$ for these transformations,
which corresponds to the orthogonal function $\varphi^\perp_0(r,{\cal E}_2) = e^{\kappa_2 r}/2\kappa_2$ and to the derivative $\frac{\partial}{\partial \mathcal{E}_2} \varphi_0(r,\mathcal{E}_2) = r e^{-\kappa_2 r}/2\kappa_2$. Equation \eref{non-trivial} can be checked in this case. Applying formula \eref{final_formula} then leads to the following potential,
valid for $\kappa_{2} \neq \kappa_{0} \neq \kappa_{1}$:
\begin{equation}
    V_4(r) = -2 \frac{\mathrm{d}^{2}}{\mathrm{d}r^{2}}\ln \mathrm{W}\left[\sinh(\kappa_{0} r), \sinh(\kappa_{1} r), e^{-\kappa_{2}r}, \beta e^{\kappa_{2}r} + r e^{-\kappa_{2}r} \right],
    \label{V4}
\end{equation}
where a global $1/2\kappa_2$ factor in the last column of the Wronskian has been removed as it disappears in the Wronskian logarithmic derivative.

\begin{figure}[h]
    \centering
    \includegraphics{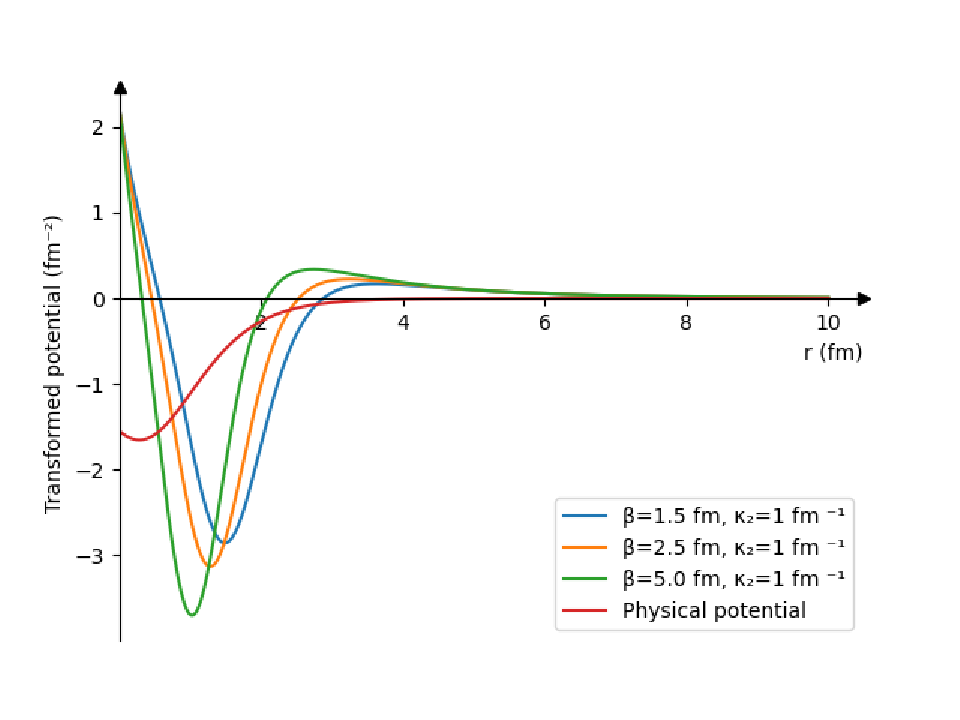}
    \caption{Phase-equivalent potentials \eref{V4} fitting the low-energy $^3S_1$ proton-neutron phase shifts obtained thanks to the supersymmetric inversion method. The physical potential \eref{V2} has $\alpha = 0$ and $\kappa_{1}=0.2315 \ \mathrm{fm}^{-1}$. }
    \label{deuteron_potential}
\end{figure}

This simple and elegant analytical potential is the most general one displaying phase shift \eref{d2}, to which $\pi$ should be added because of the bound-state addition, and one bound state.
It is the first time such a general analytical formula is found in the supersymmetric inversion scheme.
Both the energy and ANC of this bound state can be chosen arbitrarily by varying the values of $\kappa_2$ and $\beta$, respectively.
Several potentials from this family are plotted in figure \ref{deuteron_potential} for $\kappa_2=1$ fm$^{-1}$ and several values of $\beta$.
Both their phase-shift behaviour and the presence of the bound state at the expected energy were verified numerically.

Despite their mathematical interest, these potentials are non physical as their bound state does not reproduce the deuteron properties.
Reproducing the deuteron binding energy can be reached by choosing ${\cal E}_2 = -2.225$ MeV \cite{deSwart1995}.
Mathematically, and taking relativistic effects into account, this corresponds to coinciding  pseudo wave numbers $\kappa_2=\kappa_{1} = 0.2315~\mathrm{fm}^{-1}$.
In that case, already studied in \cite{baye2014}, one transformation $T_{\mathrm{r}}$ and one transformation $T_{\mathrm{l}}$ cancel each other and the potential, directly given by a non confluent pair $\{T_\mathrm{l}({\cal E}_0), T_\mathrm{add}({\cal E}_1, \alpha)\}$, reads
\begin{equation}
    V_2(r) = -2 \frac{\mathrm{d}^{2}}{\mathrm{d}r^{2}}\ln \mathrm{W}\left[\sinh(\kappa_{0} r), \alpha e^{-\kappa_{1}r} + e^{\kappa_{1}r}\right].
    \label{V2}
\end{equation}
There again, varying the value of $\alpha$ allows to arbitrarily choose the bound-state ANC, completing the whole family of phase-equivalent potentials.
The only physical potential corresponds to $\alpha=0$ and is the Eckart potential.
This potential, also displayed in figure \ref{deuteron_potential} for comparison, is the shortest-range potential of the whole phase-equivalent family: all other potentials have an asymptotic behaviour slowly decreasing as $e^{-2\kappa_1 r}$ (see \cite{baye2014} for a complete discussion of the link between this asymptotic behaviour and the bound-state ANC).

\subsection{$^{1}S_{0}$ state}

For the singlet state, we fit the phase shift on the whole elastic-scattering range.
In \cite{samsonov2002} (see also \cite{midya2015}),
a satisfactory fit was found with 6 non-confluent transformations, starting again from $V_0=0$,
with factorization parameters $\kappa_{0,\dots,5} =
0.6152, 2.0424, 4.1650, 4.6, -0.0401, -0.7540$ fm$^{-1}$.
The corresponding potential reads
\begin{eqnarray}
    V_{6}(r) & = & - 2 \frac{\mathrm{d}^{2}}{\mathrm{d}r^{2}}\ln \mathrm{W}[\sinh(\kappa_{0} r), \sinh(\kappa_{1} r), \sinh(\kappa_{2} r), \sinh(\kappa_{3} r), \nonumber \\
    && e^{-\kappa_{4}r},e^{-\kappa_{5}r}],
    \label{V6}
\end{eqnarray}
and displays again an $n_6=2$ singularity at the origin.

As the neutron-proton system does not display a physical bound state in the singlet state, adding a bound-state to this potential might seem purely academic.
Nevertheless, such {\em deep} potentials exist: they belong to the family of Moscow potentials \cite{kukulin1992} and their nonphysical bound state, also called {\em forbidden bound state}, simulates the Pauli exclusion principle between the quarks constituting the nucleons. 
It was shown in \cite{sparenberg2002} that, by exploiting the bound-state degrees of freedom (energy and ANC), such potentials were able to fit several partial waves at the same time, reducing the parity and spin dependence.
Here, this can be done by adding a bound state to potential \eref{V6} with a phase-equivalent confluent pair.
Using again formula \eref{final_formula}, this leads to potential
\begin{eqnarray}
    V_{8}(r) & = & - 2 \frac{\mathrm{d}^{2}}{\mathrm{d}r^{2}}\ln \mathrm{W}[\sinh(\kappa_{0} r), \sinh(\kappa_{1} r), \sinh(\kappa_{2} r), \sinh(\kappa_{3} r), \nonumber \\
    && e^{-\kappa_{4}r},e^{-\kappa_{5}r}, e^{-\kappa_{6}r},
    \beta e^{\kappa_{6}r} + r e^{-\kappa_{6}r}], \label{V8}
\end{eqnarray}
where parameters $\kappa_6$ and $\beta$
 could be used as free parameters to fit other partial waves.
 This is a very promising potential to revisit the ideas of \cite{sparenberg2002} is a more systematic way,
as the potentials there were built in a more phenomenological way.
However, the direct use of formula \eref{V8} is numerically unstable.
Contrary to potential \eref{V4}, neither symbolic computations, nor automatic differentiation or numerical computations were able to evaluate this expression correctly.

\section{Conclusion}

The Wronskian formulation of confluent SUSYQM transformations developed in \cite{bermudez2012, schulze2013}
was applied to the phase-equivalent transformation pairs first proposed in \cite{baye1987a, baye1987b},
replacing integral formulas for bound-state addition, removal, and modification of asymptotic normalization constant.

This allowed to unify the two steps of the SUSYQM inversion scheme proposed in \cite{sparenberg1997,baye2004} into a single Wronskian formulation,
leading to elegant analytical expressions for potentials fitting scattering phase shifts and displaying bound states.

The formalism was applied to the inversion of neutron-proton scattering phase shifts in the $^3S_1$ and $^1S_0$ waves. For the triplet state, the whole family of potentials with one bound state, reproducing the low-energy phase shift, was obtained.
Their compact analytical expression is given by equations \eref{V4} and \eref{V2} and could be tested numerically.
All these potentials have a slowly decreasing asymptotic behaviour, except for the physical potential which reproduces the deuteron bound state.

For the singlet state, the whole family of deep potentials with one Pauli forbidden state, reproducing the phase shift on the whole elastic-scattering range, was obtained.
Its analytical expression, given by equation \eref{V8}, is compact and elegant but we could not directly use it to evaluate the potential and check its properties numerically.
Indeed, the too large dimension of the Wronskian made the symbolic computation intractable. Automatic differentiation and numerical evaluation also failed.
A more robust way to evaluate this potential is thus needed before it can be used in nuclear physics calculations.
Nevertheless, the present result opens the way to a systematic search for a deep potential with minimal partial-wave dependence, as was started in \cite{sparenberg2002}.
\ack{This project has received funding from the Deutsche
Forschungsgemeinschaft within the Collaborative Research Center SFB 1245
(Projektnummer 279384907) and the PRISMA+ (Precision Physics,
Fundamental Interactions and Structure of Matter) Cluster of Excellence.
This work was supported by the Belgian Fonds de la Recherche Scientifique - FNRS under IISN Grant No.\ 4.45.10.08.
The authors thank Daniel Baye for the numerical check of equation \eref{V4}.}
\bibliography{bibliography}

\end{document}